\documentstyle[seceq,supplement,psfig,wrapft]{ptptex}

\renewcommand{\vec}[1]{\mbox{\boldmath $#1$}}

\title{
Dynamical particle-phonon couplings in proton emission from spherical nuclei
}

\author{
Kouichi {\sc Hagino}\footnote{E-mail: hagino@yukawa.kyoto-u.ac.jp}
}

\inst{
Yukawa Institute for Theoretical Physics, 
Kyoto University, Kyoto 606-8502, Japan}

\recdate{
}

\abst{
We present results of coupled-channels calculations which are performed to 
understand the role of vibrational excitations of the daughter nucleus in 
proton emission decay of spherical nuclei. 
It is shown that the 
experimental decay rates from the 2$d_{3/2}$, 3$s_{1/2}$ and
1$h_{11/2}$ states in 
proton emitting nuclei $^{160,161}$Re are reproduced simultaneously when 
the quadrupole phonon excitation of the core nuclei is taken into
account. 
We apply the formalism also to the $^{145}$Tm nucleus 
and show that good agreement is achieved 
with the recently measured branching ratio for the fine structure. 
}

\begin{document}

\maketitle

\section{Introduction}

Although nuclei beyond the proton drip line are unstable against
proton emission, their lifetime is sufficiently long to study the 
spectroscopic properties, because of the fact that 
the proton has to penetrate the Coulomb barrier. 
Thanks to the recent experimental developments of production and
detection methods, a number of ground-state as well as isomeric 
proton emitters have recently been discovered.~\cite{procon99,WD97}
These experimental data have indeed shown that the proton
radioactivity provides a powerful tool to study the structure of 
proton rich nuclei. 

In the field of heavy-ion reactions, 
the coupled-channels formalism has been a standard framework 
to extract nuclear structure information from experimental data.~\cite{HRK99} 
This formalism has been successfully 
applied also to deformed proton emitters.~\cite{ED00,KBNV00,MFL98} 
For instance, the $\beta_2$ deformation parameter of deformed 
proton emitters has been extracted from the experimental 
decay widths a/o branching ratios.  

In the previous analyses for the proton emitters in the 
region 68 $< Z < 82$, 
where proton emissions 
from the 1$h_{11/2}$, 3$s_{1/2}$, and 2$d_{3/2}$ orbitals have been
observed, the emitters have been treated as spherical nuclei without any
structure.~\cite{ASN97} 
It has been pointed out that such calculations 
systematically underestimate the measured decay
half-lives for proton emissions from the 2$d_{3/2}$ state, while the
same model works well for emissions from the 1$h_{11/2}$ and 3$s_{1/2}$ states 
in both odd-Z even-N and odd-Z odd-N nuclei.~\cite{DWB97,BBR99} 
For the $^{151}$Lu nucleus, this discrepancy 
was attributed to the 
effects of oblate deformation of the core nucleus $^{150}$Yb, which
alter both the decay dynamics and the spectroscopic factor.
~\cite{BBR99,FM00,S01} 
This fact has motivated us to investigate 
the effects of vibrational excitations on the proton decay 
of spherical nuclei.~\cite{H01,DE01} Similar effects have been well
recognized in the field of heavy-ion subbarrier fusion 
reactions.~\cite{DHRS98,S95}

In this contribution, we solve the coupled-channels equations for
spherical proton emitters in order to study 
whether the effects of vibrational 
excitations of the daughter nucleus consistently account for 
the measured decay half-life from 
the 1$h_{11/2}$, 3$s_{1/2}$, and 2$d_{3/2}$ states. 
In Sect. 2, we briefly review the coupled-channels framework for 
proton emissions. We use the Green's function approach to 
compute the decay rate of resonance states.
~\cite{ED00,ASN97,K72,BK85,BK89,DE00}
We discuss the method both for the one dimensional and for the
coupled-channels problems. 
In Sect. 3.1., we compare our results with the experimental data. 
We particularly study proton emissions from the 
1$h_{11/2}$ and 3$s_{1/2}$ states of $^{161}$Re~\cite{IDW97} and 
from the 2$d_{3/2}$ state of $^{160}$Re.~\cite{PWC92}
A clear vibrational structure in the spectrum 
of the core nucleus $^{160}$W for the proton emitter $^{161}$Re 
has recently been identified up to the $6^+_1$ state.~\cite{K01}
We shall assume that the vibrational properties are identical between the
core nuclei $^{160}$W and $^{159}$W, and study the sensitivity of 
the calculated decay rate to the dynamical deformation parameter for the 
vibration. We will also 
discuss the dependence on the multipolarity of the phonon state. 
In Sect. 3.2., we apply the formalism to the proton decay of 
the $^{145}$Tm nucleus, where the fine structure was recently
observed.~\cite{R01} 
We then summarize the paper in Sect. 4. 

\section{Green's function method for proton radioactivities}

\subsection{One dimensional case}

In this section, we review the method which we use below to calculate
the decay rate for proton emissions. Let us first consider the no
coupling case. Extension to the coupled-channels problems will be given in
the next subsection. 

We consider the following Schr\"odinger equation for a resonance state: 
\begin{equation}
\left[-\frac{\hbar^2}{2\mu}\frac{d^2}{dr^2}
+\frac{l(l+1)\hbar^2}{2\mu
r^2}+V(r)-\left(E+\frac{i}{2}\Gamma_0\right)
\right]u(r)=0,
\label{1dim}
\end{equation}
where $\mu$ is the reduced mass for the relative motion between the
valence proton and the daughter nucleus, and $V(r)$ is a potential
which consists of the nuclear, the spin-orbit, and the Coulomb parts. 
We describe the resonance state using the Gamow wave function. 
Accordingly, $u(r)$
satisfies the boundary condition that it is regular at the origin
and is proportional to the outgoing wave asymptotically, namely 
\begin{eqnarray}
u(r)&=&r^{l+1}\hspace*{4cm}~~~~~~(r\to 0), \\
&\to& {\cal N}(G_l(kr) + iF_l(kr)) \hspace*{1.5cm}~~~~ (r\to\infty),
\label{bc0}
\end{eqnarray}
where $k$ is the wave number, and $F_l$ and $G_l$ is the regular and
the irregular Coulomb wave functions, respectively. 
As is well known, these boundary conditions cannot be satisfied with a
real energy $E$, and an imaginary part of the energy is included in
Eq. (\ref{1dim}). The latter is nothing but the decay width of the
resonance state. This quantity is calculated also as a ratio between
the asymptotic outgoing flux and the normalization inside the
barrier,~\cite{MFL98}
\begin{equation}
\Gamma_0=\frac{\hbar^2k}{\mu}{\cal N}^2\left/\int^{r_t}_0\,|u(r)|^2\,dr\right.,
\label{gamma00}
\end{equation}
where $r_t$ is the outermost turning point. 

An alternative approach to calculate the decay width $\Gamma_0$ has
been proposed, which is based
on the Green's function formalism.~\cite{ED00,ASN97,K72,BK85,BK89,DE00}
This method was first developed for $\alpha$ decays by Kadmensky and 
his collaborators \cite{K72} and was recently applied to the 
coupled-channels problem for deformed proton emitters by 
Esbensen and Davids.~ \cite{ED00}

In this method, one first sets $\Gamma_0$ to be 0 in Eq. (\ref{1dim}) and 
looks for a standing wave solution which has an asymptotic form of 
\begin{equation}
\phi(r)\to {\cal \widetilde{N}}G_l(kr) \hspace*{2cm}~~~~ (r\to\infty). 
\label{stand2}
\end{equation}
The coefficient ${\cal \widetilde{N}}$ is determined so that the wave
function $\phi$ is normalized to one inside the barrier. 
Note that $\phi$ corresponds to the solution where the nuclear
phase shift is equal to $\pi$/2. 
This procedure should be a good approximation since 
$\Gamma_0$ is extremely smaller 
than the real part of the
energy $E$ (typically $\Gamma_0$ is between 10$^{-18}$ and 10$^{-22}$MeV
while $E$ is order of MeV).~\cite{WD97}
The next step is to apply the Gell-Mann-Goldberger transformation
\cite{G83} to the total Schr\"odinger equation 
$[\hat{T}-(E+i\Gamma_0/2)]\Psi=-V\Psi$, where $\hat{T}$ is the kinetic
energy operator. Adding the point Coulomb potential $V_C(r)=Z_De^2/r$,
$Z_D$ being the charge number of the daughter nucleus, to both the
left and the right hand sides, this equation is transformed to  
\begin{equation}
\left[\hat{T}+\frac{Z_De^2}{r}-\left(E+\frac{i\Gamma_0}{2}\right)\right]\Psi 
=\left(\frac{Z_De^2}{r}-V(r)\right)\Psi.
\end{equation}
The formal solution of this equation reads 
\begin{equation}
\Psi=\frac{1}{\hat{T}+\frac{Z_De^2}{r}-\left(E+\frac{i\Gamma_0}{2}\right)}
\left(\frac{Z_De^2}{r}-V(r)\right)\Psi.
\label{Green1}
\end{equation}
Next two approximations are made. Firstly one replaces $\Psi$ on the 
right hand side of Eq. (\ref{Green1}) with the standing wave solution
$\Phi$, whose radial part satisfies the boundary condition given by
Eq. (\ref{stand2}). 
Secondly, one regards $\Gamma_0/2$ as an infinitesimal
number $\eta$ which 
provides the outgoing wave boundary condition for the single particle
Green's function $(\hat{T}+V_C-E-i\eta)^{-1}$. 
As long as $\Gamma_0$ is extremely small compared with $E$, its
actual value is not important at this stage. 
As is well known, the single particle Green's
function is expressed in terms of the regular and the outgoing wave
functions as \cite{DE00}
\begin{equation}
\left\langle\vec{r}\left|
\frac{1}{\hat{T}+V_C-E-i\eta}
\right|\vec{r}'\right\rangle=
\sum_{l,m}\frac{2\mu}{k\hbar^2}\frac{O_l(kr_>)}{r_>}
\frac{F_l(kr_<)}{r_<}{\cal Y}_{jlm}(\hat{\vec{r}}_>)
{\cal Y}_{jlm}^*(\hat{\vec{r}}_<),
\end{equation}
where $O_l=G_l+iF_l$ is the outgoing Coulomb wave. 
One then obtains\cite{DE00}
\begin{eqnarray}
\Psi(\vec{r})&=&
\int\,d\vec{r}'
\left\langle\vec{r}\left|
\frac{1}{\hat{T}+V_C-E-i\eta}
\right|\vec{r}'\right\rangle
\left\langle\vec{r}'\left|
\left(\frac{Z_De^2}{r'}-V(r')\right)\right.\right.\Phi,\\
&=&
\frac{2\mu}{k\hbar^2}\frac{O_l(kr)}{r}{\cal Y}_{jlm}(\hat{r}) 
\int^{\infty}_0dr'\,F_l(kr')\left(\frac{Z_De^2}{r'}-V(r')\right)
\phi(r'),
\end{eqnarray}
as $r\to\infty$. From this equation, the normalization factor ${\cal N}$
in Eq. (\ref{bc0}) is found to be  
\begin{equation}
{\cal N}=
\frac{2\mu}{k\hbar^2}
\int^{\infty}_0dr'\,F_l(kr')\left(\frac{Z_De^2}{r'}-V(r')\right)
\phi(r').
\end{equation}
The decay rate is then obtained by Eq. (\ref{gamma00}), by 
replacing $u(r)$ with $\phi(r)$ in the denominator. 

\subsection{Coupled-channels problem}

Let us now discuss the channel coupling effects on proton
radioactivities of spherical nuclei. 
In order to take into account the vibrational core excitations, 
we consider the following Hamiltonian: 
\begin{equation}
H=-\frac{\hbar^2}{2\mu}\nabla^2 +V(r)+V_{\rm coup}(\vec{r},\alpha)
+H_{\rm vib},
\end{equation}
where 
$\alpha$ is the coordinate for the
vibrational phonon of the daughter nucleus. It is related to the
dynamical deformation parameter as 
$\alpha_{\lambda\mu}=\frac{\beta_{\lambda}}{\sqrt{2\lambda+1}} 
(a_{\lambda\mu}^{\dagger}
+(-)^{\mu}a_{\lambda\mu})$, where $\lambda$ is the multipolarity of
the vibrational mode and $a_{\lambda\mu}^{\dagger} (a_{\lambda\mu})$ 
is the creation (annihilation) operator of the phonon. 
$H_{\rm vib}=\hbar\omega \sum_{\mu}a_{\lambda\mu}^{\dagger}a_{\lambda\mu}$ 
is the Hamiltonian for the vibrational phonon. In this paper, for
simplicity, we do not consider the possibility of multi-phonon
excitations, but include only excitations to the single phonon state. 
The coupling Hamiltonian $V_{\rm coup}(\vec{r},\alpha)$ consists of three
terms, {\it i.e.,} $V_{\rm coup}(\vec{r},\alpha)=
V_{\rm coup}^{(N)}(\vec{r},\alpha)+V_{\rm coup}^{(ls)}(\vec{r},\alpha)+
V_{\rm coup}^{(C)}(\vec{r},\alpha)$. The nuclear term reads 
\begin{equation}
V_{\rm coup}^{(N)}(\vec{r},\alpha)= 
-\frac{V_0}
{1+\exp\left(\frac{r-R_0-R_0\,\alpha_{\lambda}\cdot
Y_{\lambda}(\hat{\vec{r}})}{a}\right)}-V_N(r)\,,
\end{equation}
where the dot denotes a scalar product. We have assumed that the
nuclear potential is given by the Woods-Saxon form. 
$V_N(r)$ is the nuclear component of the bare potential $V(r)$. This
term is subtracted in order to avoid the double counting. 
Notice that we do not expand the nuclear coupling potential but include the
couplings to the all orders with respect to the phonon operator.~
\cite{HRK99,HTDHL97b}
The matrix elements of this coupling Hamiltonian are evaluated using a
matrix algebra, as in Ref. \citen{HRK99}. As for the
Coulomb $V_{\rm coup}^{(C)}$ as well as the spin-orbit $V_{\rm coup}^{(ls)}$ 
terms, the effects of higher order couplings are expected to be 
small,~\cite{HTDHL97b} and we retain only the linear term. 
The Coulomb term thus reads 
\begin{eqnarray}
V_{\rm coup}^{(C)}(\vec{r},\alpha)&=& 
\frac{3Z_D e^2}{r}\,\frac{1}{2\lambda+1}
\left(\frac{R_c}{r}\right)^{\lambda}
\alpha_{\lambda}\cdot Y_{\lambda}(\hat{\vec{r}})~~~~~~({\rm for}~r > R_c) \\
&=&\frac{3Z_D e^2}{R_c}\,\frac{1}{2\lambda+1}
\left(\frac{r}{R_c}\right)^{\lambda}
\alpha_{\lambda}\cdot Y_{\lambda}(\hat{\vec{r}})~~~~~~({\rm for}~r \leq R_c), 
\end{eqnarray}
where $R_c$ is the charge radius. 
For the spin-orbit interaction, we express it in the so called Thomas 
form,~\cite{ED00,S83} 
\begin{equation}
V_{\rm coup}^{(ls)}(\vec{r},\alpha)= 
V_{\rm so}\frac{1}{r}\,\frac{df}{dr}\,\vec{l}\cdot\vec{\sigma}
+i\,V_{\rm so}R_{\rm so}\sum_{\mu}\alpha_{\lambda\mu}^*
\left\{\left(\nabla \frac{df}{dr}Y_{\lambda\mu}(\hat{\vec{r}})\right)
\cdot\left(\nabla \times \vec{\sigma}\right)\right\}\, , 
\label{ls}
\end{equation}
where $f(r)=1/[1+\exp((r-R_{\rm so})/a_{\rm so})]$. 
The last term in Eq. (\ref{ls}) 
can be decomposed to a sum of angular momentum tensors
using a formula \cite{E57}
\begin{eqnarray}
\left(\nabla g(r)Y_{\lambda\mu}(\hat{\vec{r}})\right)\cdot \vec{C}
&=&-\sqrt{\frac{\lambda+1}{2\lambda+1}}\left(\frac{dg}{dr}-
\frac{\lambda}{r}\,g(r)\right)[Y_{\lambda+1}\vec{C}]^{(\lambda\mu)}
\nonumber \\
&&+\sqrt{\frac{\lambda}{2\lambda+1}}\left(\frac{dg}{dr}+
\frac{\lambda+1}{r}\,g(r)\right)[Y_{\lambda-1}\vec{C}]^{(\lambda\mu)}.
\end{eqnarray}

In order to solve the coupled-channels equations, we expand the total
wave function as 
\begin{equation}
\Psi_{jm}(\vec{r},\alpha)=\sum_{l_pj_p}\sum_{nI}
\frac{u^{(j)}_{l_pj_pnI}(r)}{r}\,|(l_pj_pnI)jm\rangle,
\end{equation}
where 
\begin{equation}
\langle \hat{\vec{r}},\alpha |(l_pj_pnI)jm\rangle =\sum_{m_p m_I}
\langle j_p m_p I m_I|jm\rangle {\cal Y}_{j_pl_pm_p}(\hat{\vec{r}})
\varphi_{nIm_I}(\alpha), 
\end{equation}
$\varphi$ being the vibrational wave function. 
The coupled-channels equations are obtained by projecting out 
the intrinsic state $|(l_pj_pnI)jm\rangle$ from the total
Schr\"odinger equation, and read 
\begin{eqnarray}
&& \left[-\frac{\hbar^2}{2\mu}\frac{d^2}{dr^2}
+\frac{l_p(l_p+1)\hbar^2}{2\mu
r^2}+V(r)-\left(E+\frac{i}{2}\Gamma_0\right)+n\hbar\omega 
\right]u^{(j)}_{l_pj_pnI}(r) \nonumber \\
&& \hspace*{2.5cm} =
-\sum_{l_p'j_p'}\sum_{n'I'}
\langle(l_pj_pnI)jm|V_{\rm coup}
|(l_p'j_p'n'I')jm\rangle u^{(j)}_{l_p'j_p'n'I'}(r)
\end{eqnarray}
To solve these equations, one needs to compute the coupling matrix
elements of the operators 
$\alpha_{\lambda}\cdot T_{\lambda}$, where $T_{\lambda\mu}$ is either 
$Y_{\lambda\mu}$ or 
$[Y_{\lambda\pm 1}\,(-i \,\nabla \times \vec{\sigma})]^{(\lambda\mu)}$. 
These are expressed in terms of the Wigner's 6-j symbol as \cite{E57}
\begin{eqnarray}
\langle(l_p'j_p'n'I')jm|\alpha_{\lambda}\cdot T_{\lambda}
|(l_pj_pnI)jm\rangle &=&  
(-)^{j_p+I'+j}\left\{
\matrix{j & I' & j_p' \cr
\lambda & j_p & I \cr}\right\} \nonumber \\
&& \hspace*{-1cm}\times \langle {\cal Y}_{j_p'l_p'}||T_{\lambda}||
{\cal Y}_{j_pl_p}\rangle \, 
\langle \phi_{n'I'}||\alpha_{\lambda}||\phi_{nI}\rangle. 
\end{eqnarray}
For transitions between the ground and the one phonon states which we
consider in this paper, the reduced matrix element 
$\langle \phi_{n'I'}||\alpha_{\lambda}||\phi_{nI}\rangle$ is given by 
$\beta_{\lambda}$. The reduced matrix elements for the operators 
$T_{\lambda}$ are found in Ref. \citen{RG92}. 

The Green's function method discussed in the previous subsection
can provide a convenient way to calculate the decay width particularly
for the coupled-channels problems.~\cite{ED00}
The direct method seeks the Gamow wave functions 
where $u^{(j)}_{l_pj_pnI}(r)$ have the asymptotic form of 
${\cal N}^{(j)}_{l_pj_pnI}(G_{l_p}(k_{nI}r)+i\,F_{l_p}(k_{nI}r))$ 
at $r\to \infty$ for all the channels, 
where $k_{nI}=\sqrt{2\mu(E-n\hbar\omega)/\hbar^2}$ is the channel wave
number. 
This method, however, requires to solve the
coupled-channels equations in the complex energy plane and out to
large distances, which is quite time consuming and also may be
difficult to obtain accurate solutions. 
In the Green's function method, in contrast,  
the coupled-channels equations are solved in the real
energy plane and the solutions are matched to the irregular Coulomb
wave functions $G_{l_p}$ at a relatively small distance 
$r_{\rm match}$, which is
outside the range of nuclear couplings. 
From the solution of the coupled-channels equations 
$\Psi^{cc}_{jm}(\vec{r},\alpha)$ thus obtained, 
the outgoing wave function
for the resonance Gamow state is generated using the Coulomb
propagator as \cite{ED00,DE00}
\begin{eqnarray}
\Psi_{jm}(\vec{r},\alpha) &=& -\int d\vec{r}'d\alpha'
\left\langle\vec{r}\alpha \left|
\frac{1}{H_{\rm coul}+H_{\rm vib}-E-i\eta}\right|\vec{r}'\alpha'\right\rangle
\nonumber \\
&\times&\left(V(r')+V_{\rm coup}(\vec{r}',\alpha')-\frac{Z_De^2}{r'}\right)
\Psi^{cc}_{jm}(\vec{r}',\alpha'), 
\end{eqnarray}
where 
$H_{\rm coul}=-\hbar^2\nabla^2/2\mu+Z_De^2/r$ is the Hamiltonian for the
point Coulomb field. 
As in the previous subsection, the asymptotic normalization factors 
${\cal N}^{(j)}_{l_pj_pnI}$ are found to be \cite{ED00}
\begin{eqnarray}
{\cal N}^{(j)}_{l_pj_pnI}
&=&-\frac{2\mu}{\hbar^2k_{nI}}\int^{\infty}_0 dr\,r 
F_{l_p}(k_{nI}r) \nonumber \\ 
&& \times\left\langle (l_pj_pnI)jm \left|
V(r)+V_{coup}(\vec{r},\alpha)-\frac{Z_De^2}{r}\right|
\Psi^{cc}_{jm}\right\rangle, \\
&=&-\frac{2\mu}{\hbar^2k_{nI}}\int^{\infty}_0 dr\,
F_{l_p}(k_{nI}r) 
\left\{
\left(V(r)-\frac{Z_De^2}{r}\right)u_{l_pj_pnI}(r)\right.
\nonumber \\ 
&& +
\left.\sum_{l_p'j_p'}\sum_{n'I'}
\langle(l_pj_pnI)jm|V_{\rm coup}
|(l_p'j_p'n'I')jm\rangle u^{(j)}_{l_p'j_p'n'I'}(r)\right\}
\end{eqnarray}
In this way, the effects of the long range Coulomb couplings outside
the matching radius $r_{\rm match}$ are treated perturbatively. 
The partial decay width is then calculated as 
$\Gamma_{l_pj_pnI}=\hbar^2k_{nI}|{\cal N}^{(j)}_{l_pj_pnI}|^2/\mu$,
provided that the wave function $\Psi^{cc}_{jm}$ is normalized inside
the outer turning point. 

\section{Comparison with experimental data}

We now solve the coupled-channels equations and compare the results with the
experimental data. One of the most important experimental observables
is the decay half-life. In the present framework, it is calculated as 
\begin{equation}
T_{1/2}=\frac{\hbar}{S_j \Gamma_j}\ln 2,
\label{halflife}
\end{equation}
where $S_j$ is the spectroscopic factor for the resonance state, which
takes into account correlations beyond the single-particle 
picture. 
If one assumes that the ground state of an odd-Z nucleus is a
one-quasiparticle state, the spectroscopic factor $S_j$ is 
identical to the unoccupation probability for this state and is given by 
$S_j=u_j^2$ in the BCS approximation.~\cite{ASN97} 

\begin{wrapfigure}{r}{6.6cm}
\vspace*{0.8cm}
    \parbox{6.6cm}
           {\psfig{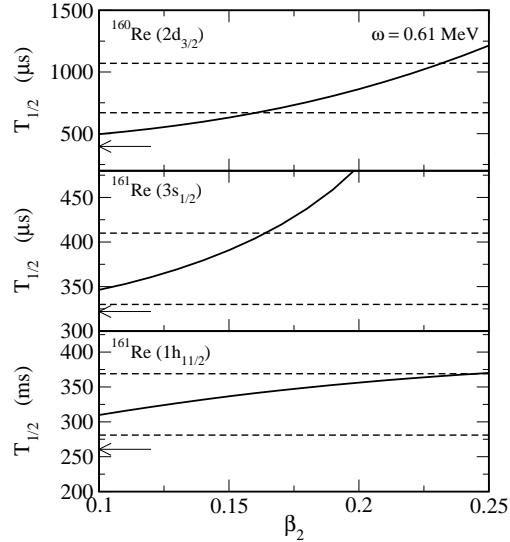}}
\protect\caption{
The decay half-lives for proton emission from the 2$d_{3/2}$ 
state of $^{160}$Re and the 
3$s_{1/2}$ and 1$h_{11/2}$ states of $^{161}$Re as a function of the 
dynamical deformation parameter $\beta_2$ of the quadrupole
vibrational excitation of the daughter nuclei. The vibrational 
excitation energy
is $\hbar\omega_2$ = 0.61 MeV. 
The experimental data are taken from Refs. 20) and 21) and
are denoted by the dashed lines. 
The arrows indicate the results in the no coupling limit.} 

\end{wrapfigure}

In all the calculations in this section, 
we use the real part of the Becchetti-Greenless optical model 
potential for the proton-daughter nucleus potential.~\cite{BG69}
The potential depth was adjusted so as to reproduce the
experimental proton decay $Q$ value for each value of the dynamical
deformation parameter $\beta$ and the excitation energy $\hbar\omega$
of the vibrational phonon excitations in the daughter nucleus. 
Following Ref. \citen{ASN97}, we assume that the depth of the spin-orbit 
potential is related to that of the central potential by  
$V_{\rm so}=-0.2V_0$. The charge radius $R_c$ is assumed to be the same as
$R_0$ in the nuclear potential. 
The spectroscopic factor $S_j$ in Eq. (\ref{halflife}) is taken from Ref. 
\citen{ASN97}. This was evaluated in the BCS approximation to 
a monopole pairing Hamiltonian for 
single-particle levels obtained with a spherical Woods-Saxon
potential. 

\subsection{$^{160,161}$Re nuclei: reconciliation of the 
$d_{3/2}$ puzzle} 

In this subsection, we solve the coupled-channels equations for 
the resonant 1$h_{11/2}$ and 3$s_{1/2}$ states in $^{161}$Re as well
as the 2$d_{3/2}$ state in $^{160}$Re. 
The first 2$^+$ state in the core nucleus $^{160}$W 
for the $^{161}$Re nucleus was recently 
observed at $\hbar\omega_2=$610 keV.~\cite{K01}Accordingly we first 
discuss the effects of the quadrupole vibrational
excitations.  
Figure 1 shows the dependence of 
the decay half-life for proton emissions from the three resonance
states on the dynamical deformation parameter $\beta_2$ 
of the quadrupole mode. 
The experimental values for the decay half-life 
are taken from Refs. \citen{IDW97,PWC92}, 
and are denoted by the dashed lines. The arrows are the half-lives in the
absence of the vibrational mode. For the decays from the 3$s_{1/2}$
and 1$h_{11/2}$ states, if one takes into account uncertainty in the
experimental $Q$ value of the proton emission, these calculations 
for the decay half-life in the no coupling limit are within the 
experimental error bars. \cite{ASN97} 
In clear contrast, the calculation for the 2$d_{3/2}$ state is still off
from the experimental data by about 30\% even when uncertainty of the
decay $Q$ value is taken into consideration.~\cite{ASN97} 

The results of the coupled-channels calculations are shown by
the solid lines in the figure. 
We treat the extra neutron as a spectator for the odd-odd proton emitter 
$^{160}$Re. 
One notices that the channel coupling
effects significantly enhance the decay half-life for the 
2$d_{3/2}$ state, while the effects are more marginal for the 
3$s_{1/2}$ and 

\begin{wrapfigure}{r}{6.6cm}
\vspace*{0.5cm}
    \parbox{6.6cm}
           {\psfig{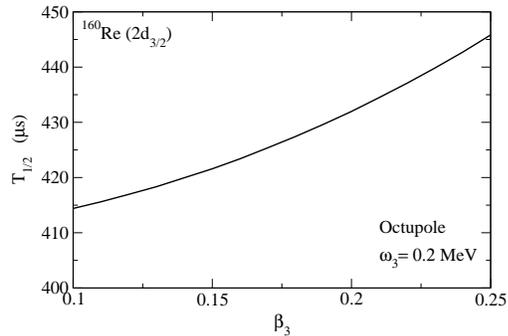}}
\protect\caption{
The decay half-life for proton emission from the 
2$d_{3/2}$ state of $^{160}$Re as a function of the dynamical
deformation parameter of the octupole phonon excitation of the core
nucleus. The excitation energy is set to be 0.2 MeV.} 
\end{wrapfigure}

\noindent
1$h_{11/2}$ states (notice the difference of the scale
of the vertical axis 
in the figure). Our results show that all of the three half-lives can
be reproduced simultaneously if the dynamical deformation parameter is
around $\beta_2\sim 0.16$. 

Why is only the $d_{3/2}$ state affected considerably by the channel
coupling effects while the others are not? A possible explanation is
as follows. The quadrupole interaction mixes the $s$ and $d$
states. For the $s_{1/2}$ state, a small admixture of $d$ states does 
not affect much, since the decay from the 
$|s_{1/2}\otimes 0^+\rangle$ state dominates after all. On the other hand, for
the $d_{3/2}$ state, a small admixture of $s$ state can influence the
decay dynamics significantly depending on the excitation energy: there 
is no centrifugal barrier for the 
$|s_{1/2}\otimes 2^+\rangle$ state, and thus the decay from this component can
compete with the decay from the dominant component 
$|d_{3/2}\otimes 0^+\rangle$ state even though the energy for the
relative motion is smaller. 
For the $h_{11/2}$ state, the quadrupole interaction mixes it with 
the $|f_{7/2}\otimes 2^+\rangle$ component. The decay from the latter may
compete with the decay from the main component due to the smaller centrifugal
potential. Compared with the $d_{3/2}$ state, however, the effective
proton energy in the incident channel 
defined as $E_p^{eff}=E_p-V_{\rm cent}(r)-V(r)$ is smaller,
which makes the channel coupling effects weaker. Therefore, the core
excitations do not influence significantly the proton decay from the 
$h_{11/2}$ state unless the coupling is very large a/o the excitation
energy is very small. 

Let us now discuss the dependence of the decay rate on the multipolarity
of the phonon mode. Figure 2 shows the effects of octupole phonon
excitations on the decay half-life from the 2$d_{3/2}$ state of 
$^{160}$Re. As an illustrative example, we take $\hbar\omega_3$=0.2
MeV, but results are qualitatively the same for different values of 
$\hbar\omega_3$. For the octupole vibration, the enhancement of the
half-life is too small to account for the observed discrepancy between
the experimental decay half-life and the prediction of the potential model
with no coupling. As was noted before,~\cite{WD97} the proton decay is
very sensitive to the angular momentum of the proton state. The same
seems to be true for the multipolarity of the collective excitation of
the daughter nucleus. 

\subsection{$^{145}$Tm nucleus: fine structure}

We next discuss the fine structure in proton emission from 
the $^{145}$Tm nucleus, which was recently observed at the Oak Ridge
National Laboratory.~\cite{R01}
In the experiment, two clear peaks were seen in the energy spectrum 
of the emitted proton, 
one corresponding to the transition from the 
1$h_{11/2}$ state in $^{145}$Tm to the
ground state 
of the daughter nucleus $^{144}$Er 
and the other to the first excited state 
at 0.326 MeV. The measured ground state decay half-life
was 3.1$\pm 0.3 \mu$s, and the branching ratio 

\begin{wrapfigure}{r}{6.6cm}
\vspace*{0.5cm}
    \parbox{6.6cm}
           {\psfig{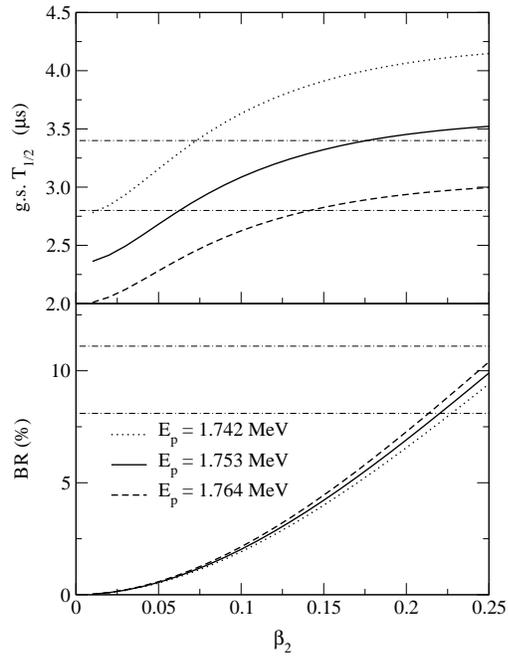}}
\protect\caption{
Sensitivity of 
the decay half-life (the upper panel) and the branching ratio (the
lower panel) for the proton emission of $^{145}$Tm 
to the dynamical
deformation parameter $\beta_2$ of the quadrupole phonon excitation of
the daughter nucleus $^{144}$Er. }
\end{wrapfigure}

\noindent
was 
9.6$\pm$1.5\%.~\cite{R01} 
Davids and Esbensen have made a detailed analysis for the fine
structure of this decay mode using the particle-vibration coupling
model.~\cite{DE01} 
They obtained good agreement with the data for both the half-life and
the branching ratio. 
Figure 3 shows results with our model Hamiltonian, which is
slightly different from the one used by Davids and Esbensen (e.g., we
include the coupling Hamiltonian for the spin-orbit interaction as
well). The experimental data are denoted by the dot-dashed lines. The
three theoretical curves are for three different proton energies,
taking its experimental uncertaintity into consideration. We find that the
ground state half-life and the branching ratio are simultaneously
reproduced if the dynamical deformation parameter $\beta_2$ is larger
than 0.22, although more studies on the dependence 
of the observables on the
strength of the spin-orbit interaction $V_{\rm so}$ 
would be needed in order to
extract a more conclusive value of $\beta_2$. Our results indicate
that the $|f_{7/2}\otimes 2^+\rangle$ state admixes 1.08\% in
probability 
with the main component 
$|h_{11/2}\otimes 0^+\rangle$ state at $\beta_2=0.18$. 
This is consistent with the
experimental suggestion given in Ref.\citen{R01}. 

\section{Summary}

We have solved the coupled-channels equations to take into
account the effects of the vibrational excitations of the spherical daughter
nucleus during the proton emission. Applying the formalism to the 
the resonant 1$h_{11/2}$ and 3$s_{1/2}$ states in $^{161}$Re and 
the 2$d_{3/2}$ state in $^{160}$Re, we have found that the
experimental data for the decay half-lives for these three states 
can be reproduced simultaneously if 
the quadrupole phonon excitation with $\beta_2 \sim 0.16$ is considered. 
This removes the discrepancy observed before between the experimental
data and the prediction of the optical potential model calculation for
the decaying 2$d_{3/2}$ state in this mass region without destroying
the agreement for the 1$h_{11/2}$ and 3$s_{1/2}$ states. 
A similar calculation with the octupole phonon was not satisfactory. 
We also discussed the fine structure in 
proton emission of $^{145}$Tm nucleus. The experimental data for the
half-life as well as the branching ratio were simultaneously reproduced
within our model, with a reasonable admixture of the excited state
components in the total wave function. 

The previous coupled-channels analyses were restricted to deformed 
proton emitters. Our studies indicate that the channel coupling
effects are significant also for spherical emitters. 
Combining with the previous results, 
our analyses show that all the observed proton
emitters can now be described in a simple single particle
picture provided that the collective motion of the daughter nucleus 
is taken into account.

\section*{Acknowledgements}
We thank Cary Davids and Krzysztof 
Rykaczewski for useful discussions.

\end{document}